\documentclass[final,1p,times]{elsarticle} 
\usepackage{graphicx} 
\usepackage{amssymb} 
\usepackage{amsthm} 
\usepackage{lineno} 

\journal{Nuclear Physics A} 
\begin{document} 

\begin{frontmatter} 


\title{Bulk Viscosity of Interacting Hadrons}

\author{A. Wiranata$^a$ and M. Prakash$^a$}

\address[a]{Dept. of Physics \& Astronomy, Ohio University, 
Athens, OH, 45701, USA}

\begin{abstract} 

We show that first approximations to the bulk viscosity $\eta_v$ are
expressible in terms of factors that depend on the sound speed $v_s$,
the enthalpy, and the interaction (elastic and inelastic) cross
section.  The explicit dependence of $\eta_v$ on the factor $\left(\frac
13 - v_s^2\right)$ is demonstrated in the Chapman-Enskog approximation
as well as the variational and relaxation time approaches.  The
interesting feature of bulk viscosity is that the dominant
contributions at a given temperature arise from particles which are
neither extremely nonrelativistic nor extremely
relativistic. Numerical results for a model binary
mixture are reported.

\end{abstract} 

\end{frontmatter} 



\section{Introduction}
Recent interest in the bulk viscosity $\eta_v$ of strongly interacting
matter stems from the observation that in the phase transition from
hadrons to strongly interacting quarks and gluons $\eta_v$
exhibits a drastic change~\cite{kapusta09}.  The purpose of this work
is to establish the dependence of $\eta_v$ on the sound speed $v_s$,
the thermodynamic properties of the system (particularly, the
enthalpy) and the interaction (elastic and inelastic) cross sections
between the constituents of a hadronic system.

\section{Bulk viscosity and the speed of sound}
\subsection{Chapman-Enskog approximation (Single component gas)}
In this approach, the first approximation to bulk viscosity can be written as
\cite{VanLeeuwen72a,prakash93}
\begin{equation}
 \eta_v = kT ~\left( \alpha_2^2 / 2w_0^{(2)} \right) \,\,\, ,
\label{bulk}
\end{equation}
where 
\begin{equation}
\alpha_2 = \frac{3}{2}\left\lbrace   z\hat{h} \left( \gamma -
\frac{5}{3}\right) + \gamma \right\rbrace\,, 
\qquad 
z=\frac{mc^2}{kT}\,,
\qquad
\hat{h} = \frac{K_3(z)}{K_2(z)}\, 
\qquad
{\rm and} \qquad 
\gamma = \frac{c_p}{c_v} \,.
\end{equation}
Above, $\hat{h}=h/c^2$ (given in terms of modified Bessel functions
for a Boltzmann gas) is the reduced enthalpy and $\gamma$ is the ratio
of specific heats at constant pressure and volume, respectively.  The
quantity $w_0^{(2)}$ is the so-called omega integral which contains
information about the cross section of the scattering
particles. Explicitly (and for illustration, for elastic scattering),
\begin{eqnarray}
w_{i}^{(s)} = \frac{2\pi z^3c}{K_2(z)^2}\int_{0}^{\infty} d\psi
\sinh^7  \psi \cosh^i\psi ~K_j(2z\cosh\psi)~~
\int_{0}^{\pi} d\Theta \sin \Theta ~\sigma(\psi,\Theta)~(1-\cos^s\Theta)~,
\label{omega1}
\end{eqnarray}
where $\sigma(\psi,\Theta)$ is the differential cross section and 
$j = \frac{5}{3}+\frac{1}{2}\left( -1\right)^i$;  the others symbols are :
\begin{eqnarray}
g &=& \frac{1}{2}(p_1-p_2)\qquad\textnormal{ and } \qquad P = 
(-p_{\alpha}p^{\alpha})^{1/2}\\
\sinh \psi &=& \frac{g}{mc}\qquad\textnormal{ and } \qquad 
\cosh \psi = \frac{P}{2mc}~.
\label{omcoeffs}
\end{eqnarray}
The adiabatic speed of sound 
$v_s = \sqrt{\left.\frac{\partial P}{\partial \varepsilon}\right |_s}$
and the ratio of specific heats $\gamma=c_p/c_v$ can be related  
using the relation 
\begin{equation}
 \gamma = 1 + \frac{\partial P }{\partial \varepsilon} = 1 + v_s^2 \,.
\label{gamma}
\end{equation}
Then, $\alpha_2$ in Eq.~(\ref{bulk}) can be rewritten in terms of the
speed of sound as
\begin{equation}
 \alpha_2 = \frac{3}{2}\left\lbrace - \left( z\hat{h} + 1 \right)
\left( \frac{1}{3}-v_s^2\right)-\frac{1}{3} z \hat{h} +\frac{4}{3}   
\right\rbrace \,.
\end{equation}
Thus, the bulk viscosity takes the form 
\begin{equation}
\eta_v = kT \frac{a^2\left(\frac{1}{3}-v_s^2 \right)^2 +2\,a\,b\,
\left(\frac{1}{3}-v_s^2 \right) +b^2}{2w_0^{(2)}}\,,
\end{equation}
where $a = -\frac{3}{2}\left( z\hat{h}+1\right)$  and  
$b = -\frac{1}{2}\left( z\hat{h}-4\right)$.

\subsection*{Limiting Situations}

It is instructive to consider the limiting cases of ultrarelativistic
and nonrelativistic situations.  For nearly massless particles or very
high temperatures $T$, $z =m/T \ll 1$.  In this case, $z\hat{h}
\longrightarrow 4 $ as $z \longrightarrow 0$.  Consequently, $a
\longrightarrow -\frac{15}{2}$ and $b\longrightarrow 0$. Utilizing
these values, the bulk viscosity can be written as
\begin{equation}
 \eta_v \rightarrow \frac{225}{8} 
\frac{kT}{2w_0^{(2)}} \left(\frac{1}{3}-v_s^2 \right)^2\,. 
\label{bulkmasslesschapmann}
\end{equation}
Notice that for weakly interacting massless particles $v_s^2
\rightarrow \frac{1}{3}$, so that $ \eta_v \rightarrow 0$. Note also
the quadratic dependence of $\eta_v$ on $v_s^2$.

On the other hand, for masses such that $z = m/T \gg 1 $, the
coefficient $ a \rightarrow -\frac{3}{2}z$ and $ b \rightarrow
-\frac{1}{2}z$. These limiting forms render the bulk viscosity as
\begin{equation}
 \eta_v = \frac{kT}{2w_0^{(2)}}\frac{z^2}{4} \left[
 9\left(\frac{1}{3}-v_s^2 \right)^2 + 6\left(\frac{1}{3}-v_s^2
 \right)+1 \right]\,. 
\label{bulkmassivechapmann} 
\end{equation} 
Weakly interacting, nonrelativistic particles are characterized by
$v_s^2 \rightarrow \frac{2}{3}$ so that, again $ \eta_v \rightarrow 0$. 

From the above analysis, we learn the important lesson that for a
given temperature and for weakly interacting particles, intermediate
mass particles contribute significantly to the bulk viscosity. It
would be interesting to investigate the extent to which this
conclusion is modified by strong interactions.

\subsection{Variational and relaxation time approximations}
Here one starts from the general
definition of the stress enegy tensor~\cite{gavin85}
\begin{equation}
T^{ij} = T^{ij}_0 + \int d\Gamma p^iv_p^j ~\delta f_p\,, \quad \delta f_p 
= -\tau \left( \frac{\partial}{\partial t} f_p^0 + v_p 
\cdot \triangledown f_p^0 \right) \quad {\rm and} \quad 
d\Gamma = \frac{d^3p}{(2\pi)^3p^0}\,,
\label{energytensor}
\end{equation}
where $f_p^0$ is the equilibrium distribution function, $v_p$ is the
velocity of a particle with momentum $p$ and energy $\epsilon_p =
p^0$. In terms of the fluid velocity field $u$, the dissipative part of the
energy stress tensor can be written as
\begin{equation}
 T^{ij}_{diss} = -\eta \left( \frac{\partial u^i}{\partial x^j} +
 \frac{\partial u^j}{\partial x^j}\right) - \left( \eta_v
 -\frac{2}{3}\right) \triangledown \cdot u \, \delta^{ij} \,.
 \label{dissipativeenergytensor} 
\end{equation} 
Inserting the deviation function
 $\delta f_p$ into the second part of the stress energy tensor and
 comparing Eqs. (\ref{energytensor}) and
 (\ref{dissipativeenergytensor}), one gets~\cite{gavin85} 
 \begin{equation}
 \eta_v = \frac{\tau}{9T}\int d\Gamma \,f^0 
\left[ \left( 1-\frac{3h}{c_vT}\right)p^2-m^2\frac{p^2}{\epsilon_p^2} 
\right] 
= \frac{\tau}{9T}\int d\Gamma \,f^0 
\left[ \left( 1-3v_s^2\right)\epsilon_p-\frac{m^2}{\epsilon_p} 
\right]^2 
\,,
\label{vary}
\end{equation}
where $\tau$ is a momentum-independent relaxation time. In writing the
rightmost equality, the identity $v_s^2=h/(c_vT)$ and other integral
identities have been used. Equation~(\ref{vary}) lends itself to
straightforward manipulations in the two limiting situations studied
in the previous section. For $m \rightarrow 0$ and $ v_s^2
\rightarrow \frac{1}{3}$, one obtains the result that $\eta_v
\rightarrow 0$.  
In the case of massive particles, for which $ \epsilon_p \approx m
$ and $ v_s^2 \rightarrow \frac{2}{3} $, the result $ \eta_v \approx
\frac{\tau m^2 n}{T} \rightarrow 0$ is obtained.  It is gratifying
that the variational approach yields results similar to those obtained
using the Chapman-Enskog approximation.

It is easy to verify that similar conclusions emerge in the case that
the expression for bulk viscosity features an energy dependent
relaxation time $\tau_a(\epsilon_a)$ \cite{kapusta09}: 
\begin{equation}
 \eta_v = \frac{1}{9T} \sum_a\,\int \frac{d^3 p}{(2\pi)^3} 
\frac{\tau_a(\epsilon_a)}{\epsilon_a^2} 
\left[ \left( 1-3v_s^2\right)\epsilon_a^2-m_a^2   \right]^2 
f_a^0 \,, 
\end{equation}
where the subscript $a$ denotes the particle species. 
In conclusion, we learn that intermediate mass particles
contribute the most to the bulk viscosity.

\section{Bulk viscosity of a binary mixture}

In the Chapman-Enskog approach, the first order approximation to the
bulk viscosity of a binary mixture has the same formal expression as
that for a one componet gas~\cite{prakash93,VanLeeuwen72b}.  Explicitly, 
\begin{equation}
\left[ \eta_v\right]_1 = kT \frac{\alpha_2^2}{a_{22}}\,, \qquad
\alpha_2 = x_1\frac{\gamma_{(1)}-\gamma}{\gamma_{(1)}-1} ~,
\end{equation}
where $x_1=n_1/n$, with $n_i$ denoting the number densities and
$\gamma_{(i)}$ denoting the ratio of specific heats of species $i$. 
The coefficient $a_{22}$ differs  from that in a one-component
gas as it receives contributions from two sources:
$a_{22}\equiv a_{22}^{'} +a_{22}^{''}~\,$. 
The quantity $a_{22}^{'}$ accounts for collisions between the same
type of particles, and is calculated from the expressions given
earlier for a one-component gas. The quantity $a_{22}^{''}$ accounts
for collisions between different types of particles. Explicitly,
\begin{equation}
a_{22}^{''} = \frac{16\,\rho_1 \rho_2}{M^2n^2} ~w_{1200}^{(1)}(\sigma_{12})~,
\label{a22}
\end{equation}
where $M = m_1 + m_2$, $\rho_i = n_i m_i$ and the appropriate omega integral
is 
\begin{eqnarray}
w_{1200}^{(1)}(\sigma_{12})& = &\frac{\pi \mu
  c^3}{4kTK_2(z_1)K_2(z_2)}
\int_0^{\infty}d\psi_{12}\sinh^3\psi_{12}
\left( \frac{g_{12}^2}{2\mu k}\right) 
\left( \frac{Mc}{P_{12}}\right)^2
~K_2\left( \frac{cP_{12}}{kT}\right)\nonumber\\
& & * \int_0^{\pi} d\Theta 
\sin\Theta_{12}\sigma_{12}(\psi_{12},\Theta_{12})(1-\cos\Theta_{12})~,
\end{eqnarray}
where
$P_{12}^2 = m_1^2c^2 +m_2^2c^2 + 2m_1m_2c^2\cosh\psi_{12}$ and
$g_{12} = m_1c \sinh\psi_{12} = m_2c \sinh\psi_{12}~$.

\begin{figure}[ht]
\centering
\includegraphics[scale=0.8]{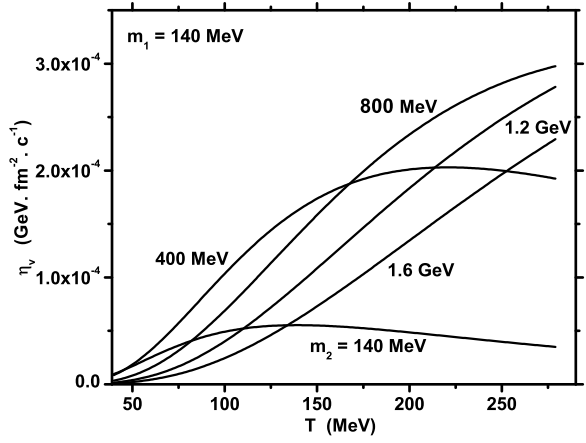}
\caption[]{Bulk viscosity $\eta_v$ versus temperature $T$ in a binary
mixture of dissimilar mass particles. The mass $m_1$ of particle 1 is fixed at 140 MeV,
whereas the mass $m_2$ of particle 2 is varied as indicated alongside the different curves. Results are for a constant
energy independent cross section of 1 fm$^2$ for all collisions.}
\label{binary}
\end{figure}

Figure~\ref{binary} shows illustrative results of bulk viscosity
versus temperature for different mass configurations with constant
cross sections.  From the results shown, we learn that 
intermediate mass configurations contribute the most to the bulk
viscosity.

Work is in progress to calculate the bulk viscosity of a hadronic
mixture comprising of many hadronic resonances whose masses extend up
to 2 GeV.

\section*{Acknowledgments} 

This research was supported by the Department of Energy 
under the grant DE-FG02-93ER40756.


\end{document}